\documentstyle[pra,aps,preprint]{revtex}
\begin{document}
\draft
\tightenlines
\title{Excitation spectrum and instability of a	two-species	
	Bose-Einstein condensate} 
\author{D.Gordon and C.M. Savage} 
\address{Department	of Physics and Theoretical Physics,	\\	 
	The	Australian National	University,	\\						  
	Australian Capital Territory 0200, Australia. \\		
	Dan.Gordon@anu.edu.au}
\date{\today}
\maketitle

\begin{abstract} 
We numerically calculate the density profile and excitation spectrum 
of a two-species Bose-Einstein condensate for the parameters of 
recent experiments.  We find that the ground state density profile of 
this system becomes unstable in certain parameter regimes, which leads 
to a phase transition to a new stable state.  This state displays 
spontaneously broken cylindrical symmetry.  This behavior is 
reflected in the excitation spectrum: as we approach the phase 
transition point, the lowest excitation frequency goes to zero, 
indicating the onset of instability in the density profile.  Following 
the phase transition, this frequency rises again.
\end{abstract}

\section{Introduction}

Following the first observation of a two-species BEC by Myatt {\it et 
al} \cite{Myatt97} there has been increased interest in the properties 
and applications of two-species Bose-Einstein condensates.  
Theoretically, two-species BECs are interesting because they allow the 
relative phase between the mean field wavefunctions to become 
manifest, whereas the phase of a single BEC cannot be observed.  The 
original experiment of Myatt {\it et al} used the 
$\left|{F,m_{f}}\right\rangle = \left|{1,-1}\right\rangle$ and 
$\left|{2,2}\right\rangle$ hyperfine sublevels of $^{87}$Rb in a 
cloverleaf trap.  The two species formed slightly overlapping clouds 
which were offset relative to one another by gravity.  Recent progress 
has also been made in the production of two-species BECs in which the 
two condensates share the same trap center, and are therefore less 
like two separate single condensates \cite{Matthews98}.  This system 
uses the $\left|{1,-1}\right\rangle$ and $\left|{2,1}\right\rangle$ 
sublevels of $^{87}$Rb which have the same magnetic dipole moment to 
first order.

In this paper, we consider the quasiparticle, or collective 
excitation, spectrum of this latter system 
\cite{Graham98,Busch97,Esry98,Pu98b}.  We find an interesting 
dependence of the excitation frequencies which clearly shows the onset 
of instability in the density profile as particle number is increased, 
and a subsequent phase transition to a stable configuration.  This 
behavior is due to the tendency of the two species to separate out 
like oil and water in certain parameter regimes \cite{Ho96}.  As in 
the two-dimensional calculation of \"{O}hberg {\it et al} 
\cite{Ohberg98}, our three-dimensional calculations show that this phase 
transition can take the form of a spontaneous breaking of cylindrical 
symmetry.  The onset of this instability as particle number is 
increased is indicated by an excitation frequency which goes to zero.  
Following a phase transition to a stable state, this frequency rises 
again to a constant value.

Our work builds on previous studies of the excitations and stability 
properties of two-species BECs.  Graham and Walls \cite{Graham98} have 
analytically solved for the excitations of a binary phase two-species 
BEC in a trap, in the Thomas-Fermi limit and in the case where the 
parameters satisfy conditions leading to a pure binary phase 
two-species BEC. Busch {\it et al} \cite{Busch97} have performed a 
variational calculation in which they compute the lowest lying 
excitation frequencies.  Esry and Greene \cite{Esry98} have 
numerically calculated the excitation spectrum of a two-species BEC in 
the in a TOP trap in which gravity separates the centers of the two 
condensates.  They found that the repulsive interaction between the 
two clouds lead to simultaneous collective excitations of both species 
which were significantly different from the single condensate case.  
However, this system did not display the stability properties 
presented here.  Pu and Bigelow \cite{Pu98a,Pu98b} have solved for the 
case of mixed Rb-Na condensate under assumptions of cylindrical 
symmetry of the density profiles, and found that this system also 
displayed instabilities of the kind presented here, however, they have 
not investigated the effect of spontaneous breaking of cylindrical 
symmetry.

\section{Ground state density profile and excitation spectrum}

For a two-species condensate, the second quantized grand canonical 
Hamiltonian is, in the position basis, \begin{eqnarray} H & = & \int 
d^{3} \hat{\mathbf{r}} \: \hat{\Psi}^{\dagger}_{A}({\mathbf{r}})\left[ 
{ \frac{\hat{\mathbf{p}}^{2}}{2m}} + V_{A}({\mathbf{r}}) - \mu_{A} 
\right]\hat{\Psi}_{A} + \int d^{3} \hat{\mathbf{r}} \: 
\hat{\Psi}^{\dagger}_{B}({\mathbf{r}})\left[ { 
\frac{\hat{\mathbf{p}}^{2}}{2m}} + V_{A}({\mathbf{r}}) - \mu_{B} 
\right]\hat{\Psi}_{B} \nonumber \\ 
& & \mbox{} + \frac{U_{AA}}{2}\int 
d^{3} \hat{\mathbf{r}} \: 
\hat{\Psi}^{\dagger}_{A}({\mathbf{r}})\hat{\Psi}^{\dagger}_{A} 
({\mathbf{r}})\hat{\Psi}_{A}({\mathbf{r}})\hat{\Psi}_{A}({\mathbf{r}}) 
\nonumber \\ 
& & \mbox{} + \frac{U_{BB}}{2}\int d^{3} \hat{\mathbf{r}} \: 
\hat{\Psi}^{\dagger}_{B}({\mathbf{r}})\hat{\Psi}^{\dagger}_{B}({\mathbf{r}}) 
\hat{\Psi}_{B}({\mathbf{r}})\hat{\Psi}_{B}({\mathbf{r}}) \nonumber \\ 
& & \mbox{} + U_{AB}\int d^{3} \hat{\mathbf{r}} \: 
\hat{\Psi}^{\dagger}_{A}({\mathbf{r}})\hat{\Psi}^{\dagger}_{B}({\mathbf{r}}) 
\hat{\Psi}_{A}({\mathbf{r}})\hat{\Psi}_{B}({\mathbf{r}})
\label{HamGC} \end{eqnarray}

where $\hat{\Psi}_{A}({\mathbf{r}})$ and 
$\hat{\Psi}_{B}({\mathbf{r}})$ are the field annihilation operators 
for species A and B respectively, $\mu_{A/B}$ are the chemical 
potentials for the two species, $m$ is the atomic mass (assumed here 
to be equal for species A and B), $V_{A/B}$ are the trapping 
potentials for the two species and the $U_{AA/BB/AB}$ are the 
scattering parameters ($U_{pq} = (4 \pi a_{pq} \hbar^{2})/m$, where 
$a_{AA/BB/AB}$ are the scattering lengths between two atoms of species 
A, two atoms of species B and an atom of species A and an atom of 
species B).

Setting $\hat{\Psi}_{A} = \psi + \hat{\delta}_{A}$ and $\hat{\Psi}_{B} 
= \psi + \hat{\delta}_{B}$, where the $\psi = 
\left\langle{\hat{\Psi}}\right\rangle$ are c-numbers and the 
$\hat{\delta}$'s contain the operator dependence, we can expand Eq.  
(\ref{HamGC}) in the deviation operators $\hat{\delta}_{A/B}$.  
Assuming that the condensate is approximately in a coherent state 
allows us to discard the third and fourth order terms in the 
$\hat{\delta}$'s,  since these are small.  The zeroth 
order term in the $\hat{\delta}$s gives the energy functional, whose 
stationary points subject to a particle number constraint are given by 
solutions to the Gross-Pitaevskii equation \cite{Lifshitz80}.  The 
first order term vanishes if the Gross-Pitaevskii equation is 
satisfied, and the second order term can be diagonalized with a 
Bogoliubov transformation, giving the quasiparticle or excitation 
energies for the BEC.
  
We use a variation of the well known basis set method \cite{Edwards96} 
to minimize the energy functional.  This method consists in working in 
a basis of harmonic oscillator eigenfunctions for the bare trapping 
potential.  We scale these basis vectors appropriately so that the 
expansion of the condensate due to repulsive atom-atom interactions is 
partially accounted for.  The scaling factors are determined by a 
variational technique using Gaussian trial wavefunctions.  This allows 
us to calculate using fewer basis vectors, or even, in certain 
geometries, to accurately scale out the dependence in one or two 
dimensions.

We define a scaled basis as follows
\begin{equation}
\phi_{i}(x,y,z) = (\lambda_{x}\lambda_{y}\lambda_{z})^{1/2}\Phi_{i}\left( 
\frac{x}{\lambda_{x}},\frac{y}{\lambda_{y}},\frac{z}{\lambda_{z}}\right)
\end{equation}
where the $\Phi_{i}(r)$ are normal modes of the free trap Hamiltonian 
and the $\lambda$'s are scaling factors determined by minimizing the 
energy functional with Gaussian trial wavefunctions.  We have found 
that we can generally achieve reasonable accuracy with around two 
hundred basis vectors, although this depends on the geometry of 
the problem and the parameter regimes explored.

In this basis,  the energy functional becomes

\begin{equation}
E = \sum_{i,j}{H^{0A}_{ij}\alpha_{i}\alpha_{j}+H^{0B}_{ij}\beta_{i}\beta_{j}}+\sum_{ijkl}\left({
\begin{array}{l} \frac{1}{2} N_{A}^{2} U_{AA} 
V^{AA}_{ijkl}\alpha_{i}\alpha_{j}\alpha_{k}\alpha_{l} \\
+ \frac{1}{2} 
N_{B}^{2} U_{BB} V^{BB}_{ijkl}\beta_{i}\beta_{j}\beta_{k}\beta_{l} \\ 
+ N_{A}N_{B}U_{AB}V^{AB}_{ijkl}\alpha_{i}\beta_{j}\alpha_{k}\beta_{l}
\end{array}
}\right)
\label{enfunct}
\end{equation}
where the $\alpha_{i}$ and $\beta_{i}$'s are the amplitudes in the 
various modes for species A and B respectively, $H^{0A/B}_{nm}$ are 
the matrix elements of the non-interacting Hamiltonian $p^{2}/2m + 
V_{A/B}(r)$ in the expanded basis and the $V^{AA/AB/BB}_{ijkl}$ are 
the matrix elements of the two body potential:
\begin{equation}
V^{pq}_{ijkl} = \int { d^{3}r 
\phi^{p}_{i}(r)\phi^{q}_{j}(r)\phi^{p}_{k}(r)\phi^{q}_{l}(r)}
\end{equation}
where p and q denote the species of atom (A or B).  The stationary 
conditions of the energy functional subject to constraints of particle 
number conservation lead to the Gross-Pitaevskii equations for a 
two-species condensate \cite{Esry97,Busch97,Pu98a}.  

We determine the ground state density profile by numerically 
minimizing the energy functional (\ref{enfunct}) subject to constant 
particle number.  We do this rather than the more usual and 
computationally simpler procedure of solving the Gross-Pitaevskii 
equation since any minimum of the energy functional is a solution to 
the Gross-Pitaevskii equation, but the set of solutions to the latter 
also include maxima and saddle points of the energy functional.  In 
the particular case considered here, solutions to the Gross-Pitaevskii 
equation can become unstable as parameters are varied, and we need to 
take this into account.  This instability is a result of the tendency 
of two-species condensates to form two separate clouds in certain 
parameter regimes.

The excitation spectrum is then calculated by solving the Bogoliubov 
equations.  In the discrete basis used here, these equations amount 
to solving the eigenproblem for a non-Hermitian matrix formed from the 
$\alpha$s and $\beta$s.  The two-species condensate case generalizes 
readily from the single condensate case, for which details can be 
found in \cite{Edwards96}.  The matrix to be diagonalized has the 
form:
\begin{equation}
\left[
\begin{array}{cccc}
P_{AA}&P_{AB}&-Q_{AA}&-Q_{AB} \\
P_{BA}&P_{BB}&-Q_{BA}&-Q_{BB} \\
Q_{AA}&Q_{AB}&-P_{AA}&-P_{AB} \\
Q_{BA}&Q_{BB}&-P_{BA}&-P_{BB} 
\end{array}
\right]
\label{Rmx}
\end{equation}
where 
\begin{eqnarray}
P^{AA}_{ij} &=& H^{0A}_{ij} - \mu_{A} + 
\sum_{k,l}{2U_{AA}V^{AA}_{ijkl}\alpha_{k}\alpha_{l} + 
U^{AB}_{ikjl}\beta_{k}\beta_{l}}
\nonumber \\
P^{AB}_{ij} &=& \sum_{kl}{U_{AB}V^{AB}_{ilkj}\alpha_{k}\beta_{l}} 
\nonumber \\
P^{BA}_{ij} &=& \sum_{kl}{U_{AB}V^{AB}_{kijl}\alpha_{k}\beta_{l}} 
\nonumber \\
Q^{AA}_{ij} &=& \sum_{kl}{U_{AA}V^{AA}_{ijkl}\alpha_{k}\alpha_{l}} 
\nonumber \\
Q^{AB}_{ij} &=& \sum_{kl}{U_{AB}V^{AB}_{ijkl}\alpha_{k}\beta_{l}}
\nonumber \\
Q^{BA}_{ij} &=& \sum_{kl}{U_{AB}V^{AB}_{jikl}\alpha_{k}\beta_{l}}
\end{eqnarray}
$P^{BB}$ and $Q^{BB}$ are defined as for $P^{AA}$ and $Q^{AA}$ 
respectively with $A \leftrightarrow B$ and $\alpha \leftrightarrow 
\beta$.  The various symmetries which exist for given trapping 
geometries and parameter regimes can be used to simplify the problem 
by separating the matrix (\ref{Rmx}) into several smaller matrices.

\section{Results}
We consider here equal trapping potentials for species A and B. This 
situation has been experimentally achieved for $^{87}$Rb atoms in the 
$\left|{F,M_{F}}\right\rangle$ = $\left|{1,-1}\right\rangle$ and 
$\left|{2,1}\right\rangle$ hyperfine sublevels which we label 
$\left|{A}\right\rangle$ and $\left|{B}\right\rangle$ 
\cite{Matthews98}.  We consider trap frequencies of $f_{x} = f_{y} = 
47/\sqrt{8}$ Hz, $f_{z} = 47$ Hz which are relevant to these 
experiments in a TOP trap.  The low inelastic collision cross section 
for these species is consistent with scattering lengths $a_{AA/BB/AB}$ 
which are all similar.  Recent calculations give $a_{AA} = 5.68$ nm 
and $a_{BB} = 5.36$ nm \cite{Burke98}.  We consider the three cases 
$a_{AB} = 5.0$ nm, 5.52 nm and 6.0 nm, which are within the range of 
expected values for this quantity, in order to show the onset of phase 
instability as $a_{AB}$ is increased.

As shown by Ho and Shenoy \cite{Ho96}, a binary condensate in the high 
particle number (Thomas-Fermi) limit can contain volumes in which only 
one species is present (giving two single particle phases, one for 
each species) and volumes in which both species coexist (binary phase) 
separated by phase boundaries.  As an example, a condensate for which 
all parameters except the interspecies scattering length are the same 
for each species will exist as a single binary phase cloud only if 
$U_{AB} < U_{AA/BB}$.  If we discard the Thomas-Fermi assumption, then 
this single species phase / binary phase picture is only approximate 
since the kinetic energy term in the Hamiltonian precludes the 
existence of sharp phase boundaries.  However, a two-species BEC can 
still undergo phase transitions in the sense that a solution to the 
Gross-Pitaevskii equation may become unstable as parameters are 
varied, and the condensate will then undergo collapse to some other 
stable solution.

FIG.  (\ref{ab1}) shows a radial cross section through the density 
profile for $a_{AB} = 5.0$ nm and $N_{A} = N_{B} = 15000$ atoms.  We 
see that in this case the two-species condensate exists as two highly 
overlapping clouds with no tendency to repel each other.  The 
corresponding excitation spectrum is shown in FIG.  (\ref{ex1} (b)), 
and FIG.  (\ref{ex1} (a)) shows a single species condensate with $a = 
\sqrt{a_{AA}a_{BB}}$.  We see that, as predicted by Graham and Walls 
\cite{Graham98}, the excitation spectrum of the two-species condensate 
undergoes a doublet splitting as compared to the case of a single 
species condensate.  We have compared our results to the high atom 
number limit derived by Graham and Walls for a spherically symmetric 
two-species condensate and found agreement to within a few percent 
with less than two hundred basis vectors.

FIG. (\ref{ex2}) shows what happens when $a_{AB}$ is increased to 
$\sqrt{a_{AA}a_{BB}} = 5.52$ nm.  We see that at high atom number, the 
lowest energy eigenvalue approaches closer to zero frequency than in 
the case of FIG. (\ref{ex1}), suggesting that we are near a region of 
phase instability in parameter space.

Indeed, in the case $a_{AB} = 6.0$ nm, the single phase solution to 
the GP equation becomes unstable at a critical atom number of around 
4000.  The minimum of the energy functional is given by a different 
density profile, as shown in FIG. (\ref{ab3}).  It can clearly be seen 
that the cylindrical symmetry is spontaneously broken by the two 
condensates' mutual repulsion.  This effect was also seen by 
\"{O}hberg and Stenholm \cite{Ohberg98} in a two dimensional 
calculation.  It illustrates the danger in assuming spherical or 
cylindrical symmetry when solving for the density profile of a binary 
condensate, since in this case such an assumption leads to a state 
which is unstable with respect to certain antisymmetric perturbations.

The excitation spectrum for the parameters of FIG. (\ref{ab3}) is shown in 
FIG. (\ref{ex3}).  We can clearly see that, for low atom numbers, the lowest 
(antisymmetric) excitation frequency goes to zero, suggesting the increasing 
instability of the condensate to antisymmetric perturbations.  This is 
indeed the case, as is shown by the breaking of cylindrical symmetry seen 
in FIG. (\ref{ab3}).  When $N \approx 4000$ is reached, the condensate 
undergoes a phase transition to an unsymmetric ground state and the lowest 
energy eigenvalue increases again to an asymptotic limit.

One further interesting feature of this excitation spectrum is the 
persistence of excitations which look like the single condensate case even 
after the phase transition is reached - for example, the nearly horizontal 
excitation at $16.6$ Hz which is hardly modified after the phase 
transition point.  We interpret this as being due to the fact that, in the 
region above the phase transition the {\it combined} density profile of the 
two species looks very much like a single species condensate, even though 
the individual density profiles of each species are greatly modified (see 
FIG. (\ref{ab3}).  The excitations in question are then interpreted as the 
normal single species type of excitations for this combined density 
profile.

\section{Conclusion}
In certain parameter regimes, two-species Bose-Einstein condensates can 
show more complex density profile behavior than single condensates.  
In particular, solutions to the Gross-Pitaevskii equation can become 
unstable or metastable as parameters are adiabatically changed, 
leading to different ground state solutions.  We have examined the way 
this effect shows up in the excitation spectra of experimentally 
realistic two-species condensates and found that the onset of instability 
is heralded by an excitation frequency which tends to zero, as would 
be expected.  The nature of the instability is consistent with the 
parity of the corresponding excitation.

\section*{Acknowledgments} The authors would like to thank Michael 
Matthews for discussions of current experiments.  Calculations were 
performed at the Australian National University Supercomputer Facility 
(ANUSF).  This work was supported by the Australian Research Council.


%
\begin{figure}
\caption{Radial cross section through the 
density profile for the parameters $a_{AB} = 5.0$ nm  and $N_{A} = 
N_{B} = 15000$ atoms.  The cross section is taken through the minimum 
of the trapping potential in the longitudinal direction.  The mesh 
plot shows the maximum density for species A and B, and the line plots 
show the densities along the x and y axes of the combined density (upper 
plots) and species A and B densities (solid and dotted lines 
respectively).  In this case, the scattering length $a_{AB}$ is small 
enough compared to the two single species scattering lengths that the 
two species show no tendency to form separate clouds.}
\label{ab1}
\end{figure}
\begin{figure}
\caption{(a) Excitation spectrum for a 
single species BEC with $a_{AB} = 5.52$ nm.  Parameters were chosen to 
approximate (b) as closely as possible.  (b) Excitation spectrum 
for the parameters of FIG.  (\ref{ab1}).}
\label{ex1}
\end{figure}
\begin{figure}
\caption{Excitation spectrum for $a_{AB} = 5.52$ nm.}
\label{ex2}
\end{figure}
\begin{figure}
\caption{Density profile for $a_{AB} = 6.0$ nm.  The top plot 
shows the densities for $N = 3000$ and the bottom plot shows the 
densities for $N = 4500$.  The spontaneous 
breaking of cylindrical symmetry is seen in the latter case.}
\label{ab3}
\end{figure}
\begin{figure}
\caption{Excitation spectrum for $a_{AB} = 
6.0$ nm.  The onset of phase instability can be seen at around $N = 
4000$ and is characterized by the lowest excitation frequency 
approaching zero.  Following a phase transition to a 
symmetry broken state, the lowest energy eigenvalue increases again, 
tending towards an asymptotic limit.}
\label{ex3}
\end{figure}

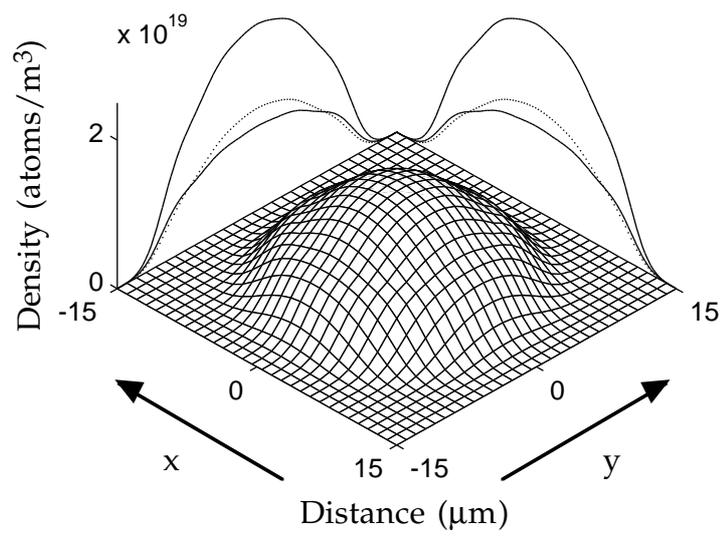
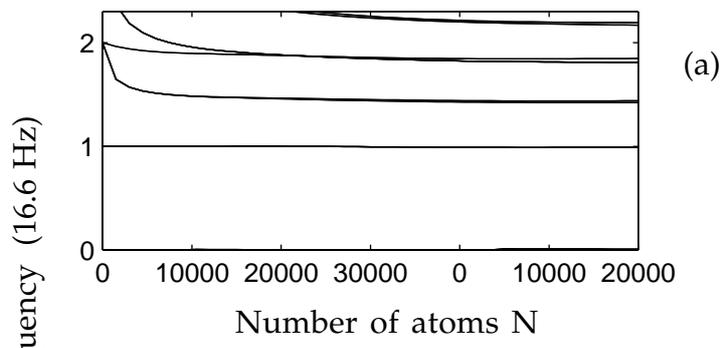
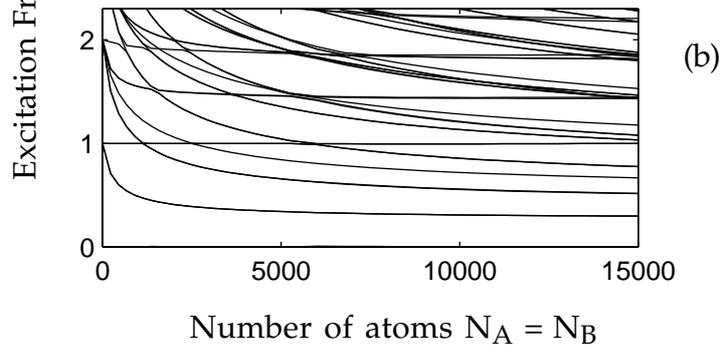
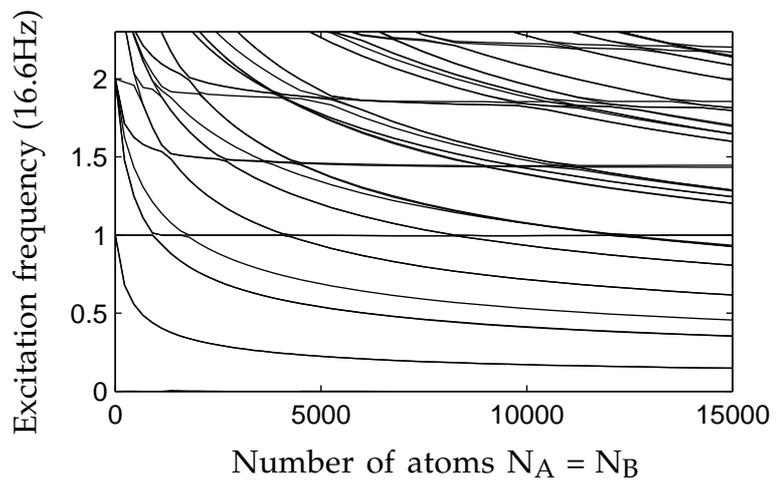
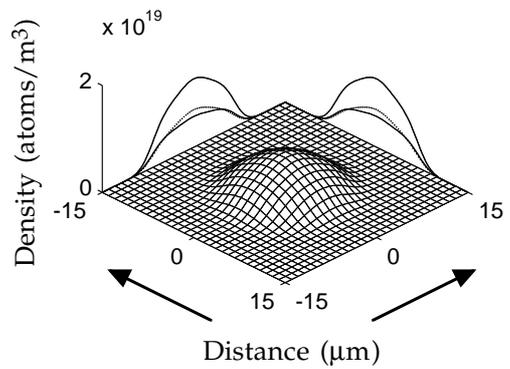
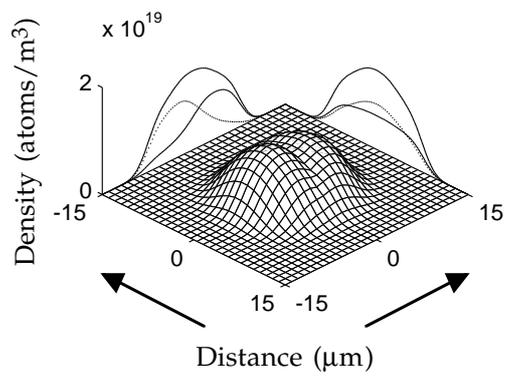
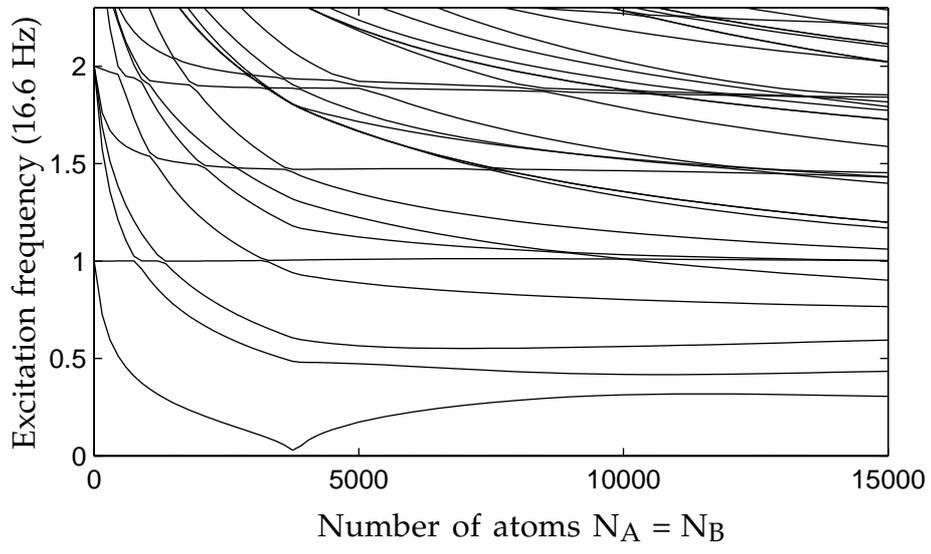
\end{document}